\documentclass[extras,11pt]{article}

\usepackage{fullpage,ltexpprt, amsmath, amssymb, epsfig, pxfonts, algorithm, algorithmic, color, tikz}
\usepackage{hyperref}
\usetikzlibrary{%
decorations.fractals,%
decorations.shapes,%
decorations.text,%
decorations.pathmorphing,%
decorations.pathreplacing,%
decorations.footprints,%
decorations.markings}

\newcommand{\INPUT}{\item[{\bf Input:}]}
\newcommand{\OUTPUT}{\item[{\bf Output:}]}
\def \HK{{\textrm{OPT}_{\textrm{HK}}}}
\def \genus {\gamma}
\def \girth {g}
\def\surface {\Sigma}

\def\Pointsize {1.4pt}
\def\mingirth{\alpha(\genus)}

\begin{document}

\title{\Large The Asymmetric Traveling Salesman Problem   on Graphs with Bounded Genus}

\author{
Shayan Oveis Gharan%
    \thanks{ Department of Management Science and Engineering, Stanford University, Stanford, CA 94305. Email:\protect\url{{shayan,saberi}@stanford.edu}.}
\and
  Amin Saberi\footnotemark[1]%
}

\date{}

\maketitle

\pagenumbering{arabic}
\setcounter{page}{1}

\begin{abstract}
\small\baselineskip=9pt
We give a constant factor approximation algorithm for the asymmetric
traveling salesman problem when the support graph of the solution of the
Held-Karp linear programming relaxation has bounded orientable genus.
\end{abstract}

\section{Introduction}

We present the first constant-factor approximation algorithm for the
Asymmetric Traveling Salesman Problem (ATSP) for metrics defined by a
weighted directed graph with a bounded
orientable genus. This is a natural special case: consider a
metric obtained by shortest path distances in a city with one way
streets and a constant number of bridges and underpasses.

The result  is more general: we can obtain constant factor
approximation algorithms when the underlying graph of the fractional
solution of the Held-Karp linear programming relaxation has bounded orientable
genus. It is easy to see that this
is a less strict condition. In fact, it is known that the corner
points of the Held-Karp relaxation polytope define very sparse
graphs \cite{goemans06} and  in practice they often turn out to be
planar.

The symmetric version of this problem (STSP) has been studied
extensively on Euclidean~\cite{aroraeuclidean}, planar
\cite{Papadimitriou:planartsp,Arora:planartsp,kleintsp}
or low-genus metrics \cite{Demaine:boundedgenus}. But to the best of our knowledge, this is the
first result of this type for ATSP.

Our algorithm  rounds the solution of the Held-Karp
linear programming  relaxation.
Therefore, it also gives a constant
upper bound on the integrality gap. It is worth noting that the
best-known constructions that lower bound the integrality
gap~\cite{charikar:integralitygap}  are also planar.

%


Our result  builds on a central lemma in Asadpour et al. \cite{atsploglogn} that shows for finding a
constant-factor approximation algorithm for ATSP, it is sufficient to
find a ``thin'' tree
in the fractional solution. Roughly speaking, a tree is $\epsilon$-thin with
respect to a graph $G$, if it does not contain more than an
$\epsilon$-fraction of the edges of $G$ across any cut.

On the other hand, we use a different approach for proving thinness: we use the
the embedding of the graph and its geometric dual. In particular, we take advantage
of the correspondence between the cutsets of the graph $G$ and cycles
of the dual graph $G^*$.  If $G^*$ does not have any short cycles and
all the edges of $T$ are far apart in $G^*$, then $T$ can not contain
too many edges from any cut of $G$ and therefore it is thin.

Thin trees were first defined in the graph embedding literature in an attempt to
prove a notoriously difficult conjecture by Jaeger on the existence of
certain nowhere-zero flows~\cite{goddyn:thintreeconj}. Our result on
the existence of thin trees in graphs with bounded genus also proves a
weaker version of Jaeger's conjecture and essentially implies the main
result of  \cite{zhang:jaeger}.  Furthermore, our algorithm and its
proof is simpler than the vertex splitting argument of
\cite{zhang:jaeger} that uses case analysis.

We were informed recently \cite{c10} of an application of the
result of this paper for the {\em minimum stabbing spanning tree
problem} on the plane. The problem is as follows: we are given a set
$P ={p_1, \ldots, p_n}$ of points in $R^2$. The task is to construct a
spanning tree on $P$ by connecting vertices with straight lines such
that the crossing number, which is the maximum number of segments that
are encountered by any line, is minimized. Fekete et al \cite{flm04}
show that the natural linear programming relaxation of the problem
contains a fractional optimal solution whose support is planar. This
has a very interesting implication: the result of our paper on the
existence of  thin trees in planar graphs also gives a constant factor
approximation for this problem! The best previously known
approximation algorithm   was $O(\log n/\log\log n)$ by Chekuri et
al. \cite{cvz09} improving over  Bilo et al. \cite{bgrs04} and
Har-Peled \cite{h09}.  For a more detailed discussion, see
Appendix \ref{sec:stabbingtree}.

In the rest of this section, let us sketch the main steps of the
algorithm and its analysis.
The  main result of the paper is to find a polynomial-time algorithm
for finding an $f(\genus)/k$-thin tree in a $k$-edge connected graph
of genus $\genus$. In order to do that, we
first show how to find such a tree if the dual of the graph has high girth.
This is sufficient for planar graphs.  The dual of every cutset in $G$
is a cycle
in $G^*$. Therefore if $G$ is $k$-edge connected, $G^*$ has
girth at least $k$.
In graphs with genus even slightly bigger than zero, high connectivity
does not imply high dual girth. In fact, the dual
of a graph with high edge  connectivity can have several short cycles.
In section
\ref{sec:dual-girth}, we show how to remove short cycles from $G^*$
without creating too many connected components in $G$. For doing
this, we have to use fairly simple ``surgical'' operations like
cutting handles and adding
topological disks to the surface. The reader interested in the
algorithm for planar graphs can
skip this section and read sections 2,3, and 5.

In the last section,  we make the connection  between ATSP and thin
trees concrete. We show that an algorithm that
finds an $O(1)/k$-thin spanning tree in a $k$-edge connected graph gives a
constant factor approximation algorithm for ATSP.

\section{Preliminaries}
\label{sec:preliminaries}

In the Asymmetric Traveling Salesman problem (ATSP),  we are given a
set $V$ of $n$ points and a cost function  $c: V \times V
\rightarrow \mathbb{R}^+$. The goal is to find the minimum cost tour
that visits every vertex at least once. Since we can replace every
arc $(u,v)$ in the tour with the shortest path from $u$ to $v$, we
can assume $c$ satisfies the triangle inequality.

Recently, Asadpour et al. \cite{atsploglogn} obtained an $O(\log{n}/\log{\log{n}})$-approximation algorithm for this problem
improving the results of~\cite{atsplogn,Blaser02,Kaplan05,FeigeS07}. For the symmetric version, the $3/2$-approximation
algorithms by Christofides~\cite{Christofides76} is still the best known.

Given an instance of ATSP corresponding to the cost function $c: V \times V \rightarrow \mathbb{R}^+$, we can obtain a lower bound on the optimum value by considering the following linear programming relaxation defined on the complete bidirected graph with vertex set $V$:
\begin{eqnarray}
\label{atsplp}
\textrm{min} & & \sum_{a} c(a) x_a \\
\textrm{s.t.} & & x(\delta^+(S)) \geq 1 ~~~~~~~~~~~~~~~~~\forall S \subset V,  \label{eq:d+} \\
& & x(\delta^+(v)) = x(\delta^-(v)) =1 ~~~~\forall v \in V, \label{eq:eul}\nonumber \\
& & x_a \geq 0  ~~~~~~~~~~~~~~~~~~~~~~~~~  \forall a \nonumber.
\end{eqnarray}

In the above linear program $\delta^+(S)$ ($\delta^-(S)$) denotes the set of directed edges leaving (entering) $S$ in the bidirectional complete graph on $V$.
This linear programming relaxation is known as the Held-Karp
relaxation \cite{HeldKarp70}, and its optimum value, which we denote
by $\HK$, can be computed in polynomial time.

The focus of this paper is on the special case in which $c$ is the distance function
in a directed planar (or bounded-genus) graph $G$. Even when $G$ is planar, the optimal solution of the LP relaxation is not necessarily a planar graph. However, it
is easy to obtain a planar solution of the same cost, by replacing every edge $e \notin G$ with the edges of the shortest path connecting its endpoints in $G$. The
same argument works for graphs with bounded genus.

Let $x$ be a feasible  solution of Held-Karp relaxation
(\ref{atsplp}). Our approximation algorithm will be based on
rounding $x$ to an integral solution.  For this rounding, it turns
out that it is sufficient to find a spanning tree that is thin with
respect to the appropriately defined multi-graph representing $x$. In the rest of the paper, we refer to multi-graphs (graphs with loops and
parallel edges) simply as graphs.

\begin{definition}
A subset $F\subseteq E$  is $\alpha$-thin with
respect to $G$, if for each set $U\subset V$,
$$|F(U,\overline{U})|\leq \alpha |E(U,\overline{U})|,$$
where $F(U,\overline{U})$ and $E(U,\overline{U})$ are respectively
the sets of edges of $F$ and $E$ that are in the cut  $(U,
\overline{U})$.
\end{definition}


\subsection{Surfaces and graph embedding}
We also need to recall some of the concepts in topological graph theory.
By a {\em surface}, we mean a compact connected 2-manifold without boundary.
It is well known that all surfaces are classified into the sphere with $\genus$ handles (denoted by $S_\genus$) or the crosscaps (denoted by $N_\genus$) and are called orientable and non-orientable surfaces respectively.
 Throughout
this paper by a surface we mean an orientable surface, and all the
theorems have been proved for orientable surfaces.

In the above definition, $\genus$ represents the genus of the surface. An equivalent definition for the genus of an orientable surface is the maximum  number of disjoint simple closed curve which can be cut
from the orientable surface without disconnecting it.

An embedding  of a  graph $G$ into a surface $\surface$, is a
homeomorphism $i:G\rightarrow \surface$ of G into $\surface$. 
The {\em orientable  genus} of a graph $G$ is the minimum $\genus$
such that $G$ has an embedding in $S_\genus$. For
example, planar graphs have genus zero.

Let $G(V,E)$ be a  graph embedded on a surface $\surface$. A set
$S\subseteq E$ on $\surface$ is {\em separating} if $\surface-S$ is
disconnected; otherwise $S$ is called {\em non-separating}. For
instance, the definition of orientable genus implies that any set of
$\genus(\Sigma)+1$ disjoint cycles of $G$ is separating.

Suppose that we have  embedded $G$ on a surface $\Sigma$. The {\em
geometric dual} of $G$ on $\Sigma$, $G^*$, is defined similar to the
planar graphs. Particularly, The vertices of $G^*$ correspond to the
faces of $G$. The edges of $G^*$ are in bijective correspondence $e
\rightarrow e^*$ with the edges of $G$, and the edge $e^*$ joins the
vertices corresponding to the faces containing $e$ in $G$.
For a more
extensive discussion of embeddings of graphs in surfaces, see
\cite{moharthomassen}.


\section{Constructing a thin-tree}
\label{sec:findthintree}


Let $G(V,E)$ be a connected graph embedded on an orientable surface,
and $G^*$ be its geometric dual. The dual-girth of $G$, denoted by
$g^*(G)$ is the length of the shortest cycle in $G^*$.
The main result of this section is the following lemma.

\begin{lemma}\label{lem:dual-girth} A connected graph embedded on an orientable surface
with genus $\gamma$ and dual-girth $g^*$ has a spanning tree with thinness
$\frac{2\mingirth}{g^*}$, where $\mingirth =4+\lfloor
2\log_2{(\genus+\frac{3}{2})}\rfloor$. Furthermore, such a tree can
be found in polynomial time.
\end{lemma}

We will prove this lemma in the rest of this section.
First note that
if $g^* =1$, the lemma holds for trivial reasons.
Therefore, without loss of generality assume that $g^*  > 1$. That implies
that no face of $G$ can have two copies of an edge.
In particular,  $G$ does not have any cut edge.

Define the distance of two edges in a graph to be the closest
distance between their endpoints.  Our most basic tool for
establishing the thinness of a tree $T$ in $G$ is to relate it to
the pairwise distance of its corresponding edges $T^*$ in $G^*$. If
$G^*$ does not have any short cycles and all the edges of $T^*$ are
far apart in $G^*$, then the tree can not contain too many edges
from any cut. We will establish that for any subset of
edges:

\begin{lemma}
\label{lem:thinnesstest} Let $F$ be a set of edges in $G$ and $F^*$
be the corresponding edges in the dual. If for some $m \leq g^*(G)$,
the distance between each pair of edges in $F^*$ is at least $m$,
then $F$ is   $\frac{1}{m}$-thin in $G$.
\end{lemma}
\begin{proof}
Consider a cut $S=(U,\overline{U})$ in $G$. Let us start by
showing that $S^*$ is a collection of edge-disjoint cycles
$C_1,C_2,\ldots,C_l$ in $G^*$. This is because the number of edges from $S^*$ incident to a
vertex in $G^*$ is equal to the intersection of
$S$ with its corresponding face in $G$ and that is an even number. Otherwise,
either that face contains two copies of an edge of $S$, or
one could find a path $P$ in that face such that $P\cap
S=\emptyset$, while the endpoints of $P$ are in different sides of
the cut, which are both impossible.

Because the distance of each pair of edges in $F^*$ is at least $m$,
 $F^*$
can not have more than $\max(1,\lfloor length(C_i)/m \rfloor)$ edges
in $C_i$, for $1\leq i\leq l$. Therefore,
$$|F^*| \leq \sum_{i=1}^l \max(1,\lfloor \frac{length(C_i)}{m}\rfloor)= \sum_{i=1}^l \lfloor \frac{length(C_i)}{m}\rfloor \leq \frac{|S^*|}{m}.$$
Note that the equality holds by the assumption $length(C_i)\geq
g^*\geq m$. Thus the number of edges of $F$ in the cut
$(U,\overline{U})$ is no more than $\lfloor
|(U,\overline{U})|/m\rfloor$ and $F$ is  $1/m$-thin.
\end{proof}

Considering the above Lemma, our goal will be to find a set of edges
in $G^*$ that are sufficiently far apart. We will do this by finding long
threads iteratively and selecting one edge from each thread.

A {\em thread} in a graph $G$ is a maximal subgraph of $G$ which is
\begin{itemize}
\item a path whose internal vertices all have degree $2$ in $G$
and its endpoints have degree at least 2, or
\item a cycle in which
all vertices except possibly one have degree $2$.
\end{itemize}

\begin{algorithm}
\caption{Finds a thin tree in a graph with large dual-girth}
\label{alg:computethinset}
\begin{algorithmic}[1]
\INPUT A connected graph $G$ embedded on an orientable surface with genus $\genus$, and its dual $G^*$ with girth $g^*$.
\OUTPUT A spanning tree $T$ with thinness at most
$\girth^*/2\mingirth$. \STATE $F^* \leftarrow \emptyset$
\WHILE {there exists an edge in $G^*$}
\STATE Find a thread $P$ of length at least $\girth^*/\mingirth$ in
$G^*$. \label{alg:step:threadselection}
\STATE Add the middle edge of $P$ to $F^*$ and remove it from
$G^*$. If  $P$ is a cycle, define its middle edge to be the one with the maximum distance from the high-degree vertex.
\STATE \label{algstep:degreeonedeletion} Iteratively delete
all the degree one vertices with their incident edges. \ENDWHILE
\RETURN A spanning tree $T \subseteq F$, where $F$ is the set of
edges corresponding to $F^*$ in $G$.
\end{algorithmic}
\end{algorithm}

 Let us start by showing the existence of long threads. That is a
straightforward application of the result of Goddyn et al. \cite{goddyn:largegirth}.

\begin{lemma}
\label{lem:longthreadexist} A graph  with minimum degree $2$ and
girth $\girth$, embedded on a surface with genus $\genus$ has a
thread of length at least $\girth/\mingirth$, where $\mingirth =4+\lfloor
2\log_2{(\genus+\frac{3}{2})}\rfloor$.
\end{lemma}

\begin{proof}
Let $H$ be a graph satisfying the conditions of the theorem and $H'$
be the graph obtained by iteratively replacing the vertices of
degree $2$ in $H$ with an edge.
In other words, let $H'$ be the graph obtained by replacing every thread
in $H$ by an edge. By using the following result of  Goddyn et al.\cite{goddyn:largegirth}, we know that the girth of $H'$ is at most $\mingirth$:
\begin{theorem}[Goddyn et al. \cite{goddyn:largegirth}]
Let $k_\lambda$ denote the least integer such that all graphs with genus at most $\lambda$, and minimum degree at least 3 have girth at most $k_\lambda$. Then, for any $\lambda\geq 0$, we have:
$$
k_\lambda \leq 4+\lfloor 2+\log_2(\gamma+3/2) \rfloor
$$
\end{theorem}
Therefore $H'$ has a cycle of length at most $\mingirth$. Now it is
easy to see that at least one of the edges of that cycle is obtained
from a thread of length at least $\frac{\girth}{\mingirth}$ in $H$.
\end{proof}

Because of the above lemma, Algorithm \ref{alg:computethinset} terminates in polynomial time.
The algorithm has an equivalent description in terms of the original
graph $G$. Roughly speaking, in each iteration, we find a collection
of consecutive parallel edges,  add the middle edge from that
collection to $F$ and contract the end points. The
embedding is crucial for the execution of this procedure because it
provides a notion of a middle edge, and the notion of consecutive
parallel edges (parallel edges that form a face).
%

It is also worth noting that $|F|$  may end up being bigger than $|V(G)| - 1$ in an execution of Algorithm  \ref{alg:computethinset}. This is because a thread in $G^*$ may be equivalent to a  collection of parallel {\em loops}.
The next lemma immediately proves Lemma \ref{lem:dual-girth}.

\begin{lemma}
\label{lem:faredgeselection}  The set $F$ computed in Algorithm
\ref{alg:computethinset} is connected and spanning in $G$.
Furthermore, the pairwise distance of the edges of $F^*$ in $G^*$ is
at least $\girth^*/2\mingirth$.
\end{lemma}
\begin{proof}
For the proof of the first statement, consider a non-empty cut
$S=(U,\overline{U})$ in $G$, and let $S^*$ be its dual. As we argued
in the proof of Lemma \ref{lem:thinnesstest},  $S^*$ is a collection
of cycles. It is also easy to see that Algorithm
\ref{alg:computethinset} selects at least one edge from each cycle of $G^*$, thus at least one edge  from $S$. For any cycle $C$ in $G^*$, the first thread $P$ selected in step \ref{alg:step:threadselection} of Algorithm \ref{alg:computethinset} that has a non-empty intersection
with $C$ should lie completely in $C$ (i.e. $P\subset C$). Therefore the middle edge of $P$,  added to $F^*$ is certainly an edge of $C$.

For the second statement, first observe that after adding an edge
$e$ to $F^*$, the algorithm immediately removes all the edges that
are of distance less than $g^*/2\mingirth$ from $e$. This is because
all these edges are a part of the thread and therefore they are
deleted sequentially.

Furthermore, although each iteration of the while loop may  increase the
distance of some pairs of edges, it never increases the distance of
two edges that are closer than $g^*/\mingirth$. Therefore, the distance
of any pairs of edges that are closer than $g^*/ \mingirth$ remains the
same until one of them is deleted.
\end{proof}


\section{Increasing the girth of dual graph}
\label{sec:dual-girth}
Let  $G(V,E)$ be a planar graph and $G^*$ be its geometric
dual.
In the
previous section we showed that the girth of $G^*$ plays an important role in finding
a thin spanning tree in $G$.
By Whitney's theorem \cite{whitney:planargraphs}, $S\subseteq E$ is a cutset
(minimal cut) in a {\em planar} graph $G$ if and only if $S^*$ is a cycle in $G^*$.
Therefore, if $G$ is planar and $k$-edge connected, the girth of $G^*$ will be at
least $k$.  Unfortunately, this relation does not hold for non-planar graphs as their dual may have very small cycles.

In the rest of this section we show that we can get rid of these small cycles by deleting their edges. Deleting these edges may result in making the graph disconnected. We will try to increase the girth as much as possible while creating only a small number of connected components.

Later, we will find a thin tree in every connected component and merge them into a spanning tree using an arbitrary set of edges from the original graph. Since the number of connected components is small, this is possible with only a small loss in the thinness of the final spanning tree. In fact, in the statement of Theorem \ref{thm:constantthintree}, the $O(\sqrt{\genus})$ dependence of the thinness on the genus of the surface comes from the   balance between the number of connected components at the end of the procedure with the girth of the final graph in Lemma \ref{lem:smallconlargegirth}.

\begin{lemma}
\label{lem:smallconlargegirth}
Let $G$ be a $k$-edge connected graph embedded on an orientable surface with genus $\genus > 0$, and $G^*$ its geometric dual. There is a polynomial time algorithm that deletes some of the edges of $G$, and obtains a new graph $H$ with the dual $H^*$ such that $H$ has  at most $2\sqrt{\genus}$ connected components while $girth(H^*) \geq \frac{k}{3\sqrt{\genus}}$.
\end{lemma}
The algorithm considers each small cycle in $G^*$, and simply deletes its corresponding edges from $G$, and updates $G^*$ accordingly.
\begin{algorithm}
\caption{Constructing a high girth dual}
\label{alg:incdualgirth}
\begin{algorithmic}[1]
\INPUT A $k$-edge connected graph $G$ embedded on a surface with genus $\genus > 0$ and its dual $G^*$.
\OUTPUT A graph $H$ and its dual $H^*$ where $\kappa(H) \leq 2\sqrt{\genus}$ and $girth(H^*)\geq \frac{k}{3\sqrt{\genus}}$.
\WHILE {$girth(G^*) < \frac{k}{3\sqrt{\genus}}$}
\STATE Find a
cycle $C^*$ of length less than $\frac{k}{3\sqrt{\genus}}$ in $G^*$.
\STATE
Delete its corresponding edges from $G$, and update $G^*$ accordingly.
\ENDWHILE
\RETURN $G$ and $G^*$
\end{algorithmic}
\end{algorithm}

To see the effect of this cycle deletion process we use the
following lemma.

\begin{lemma}
\label{lem:cycleremovalproperty} Let $G$ be a non-planar graph
embedded on an orientable surface $\Sigma$ with genus $\genus$, and
$G^*$ its geometric dual. If $C$ is the set of corresponding edges
of a cycle $C^*$ in $G^*$, then either $G-C$ can be embedded on a
surface with smaller genus, or $\kappa(G-C) > \kappa(G)$, where
$\kappa(G)$ is the number of connected components of $G$.
\end{lemma}

\begin{proof}
We define a surgery operation in which the surface $\Sigma$ is cut
along the simple curve defined by $C^*$,  and then  a topological
disk is attached to each side of the cut. We will show how $G-C$ can
be embedded on the resulting surface (or surfaces).

If $C^*$ is a non-separating cycle in $G$, then cutting along $C^*$
and adding the two topological disks removes one of the handles of
$\Sigma$, thus giving rise to a unique connected surface with
smaller genus. The edges of $C$ crossing the curve $C^*$ are removed
from $G$, therefore $G - C$ is embeddable on the new surface. Figure
\ref{fig:cyclecutting}, shows the details of this operation.

\begin{figure*}[hta]
\centering

%



\begin{tikzpicture}[scale=.6]
\draw [fill,color=black!8!white!92] (-5.5,3) rectangle (3.5,-2);
\draw [fill,color=black!8!white!92,decorate,decoration={zigzag,amplitude=1.5mm,segment length=5mm,post length=0mm}] (-5.7,2.8) rectangle (-.7,-1.7) (1.7,2.8) rectangle (3.7,-1.8);
\draw [line width=1.2pt](-5.5,3) -- (3.5,3) (-5.5,-2) -- (3.5,-2);


\draw [color=red] (0,3) arc (90:270:.5cm and 2.5cm);
\draw [color=red,dashed] (0,-2) arc (-90:90:.5cm and 2.5cm);

tikzstyle{every node}=[shape=circle,color=red,fill=red,inner sep=\Pointsize];
\draw [color=red,fill]  (0,.5) +(185:.5cm and 2.5cm)  circle (.12cm);
\draw [color=red,fill]  (0,.5) +(125:.5cm and 2.5cm)  circle (.12cm);
\draw [color=red,fill]  (0,.5) +(153:.5cm and 2.5cm)  circle (.12cm);
\draw [color=red,fill]  (0,.5) +(230:.5cm and 2.5cm)  circle (.12cm);

\tikzstyle{every node}=[draw,shape=circle,fill=black,inner sep=\Pointsize];

\path (-2.5,0) node (v0) {$$};
\path (v0) +(.5,2) node (v1) {$$};
\path (v0) +(0,-1.5) node (v2) {$$};
\path (v1) +(4,0) node (v3) {$$} edge (v1);
\path (v0) +(4.5,0) node (v4) {$$} edge (v3) edge (v1) edge (v2);
\path [dashed] (v2) +(4.5,0) node (v5) {$$} edge (v0);
\path (v0) +(-1,0) node (v6) {$$} edge (v1) edge (v2);

\draw [color=black] (v6)-- +(-1,0) (v3)-- +(1,0) (v4)-- +(1,0);
\draw [dashed] (v0)-- +(-2,-1.5) (v5)-- +(1,0)  (v5) -- +(1,1);
\draw (v6) arc (180:90:.5cm and 3cm);
\draw [dashed] (v0) arc (0:90:.5cm and 3cm);

\draw [->,>=latex,line width=3.5pt] (4,0) -- (6,0);

\begin{scope}[xshift=12cm]
\draw [fill,color=black!8!white!92] (-5.5,3) rectangle (-1,-2) (1,3) rectangle (3.5,-2);
\draw [fill,color=black!8!white!92,decorate,decoration={zigzag,amplitude=1.5mm,segment length=5mm,post length=0mm}] (-5.7,2.8) rectangle (-.7,-1.7) (1.7,2.8) rectangle (3.7,-1.8);
\draw [line width=1.2pt] (-5.5,3) -- (-.8,3) (-5.5,-2) -- (-.8,-2) (.8,3)--(3.5,3) (.8,-2)--(3.5,-2);

\tikzstyle{every node}=[draw,shape=circle,fill=black,inner sep=\Pointsize];
\path (-2.5,0) node (v0) {$$};
\path (v0) +(.5,2) node (v1) {$$};
\path (v0) +(0,-1.5) node (v2) {$$};
\path (v1) +(4,0) node (v3) {$$} ;
\path (v0) +(4.5,0) node (v4) {$$} edge (v3) ;
\path [dashed] (v2) +(4.5,0) node (v5) {$$} ;
\path (v0) +(-1,0) node (v6) {$$} edge (v1) edge (v2);

\draw [color=black] (v6)-- +(-1,0) (v3)-- +(1,0) (v4)-- +(1,0);
\draw [dashed] (v0)-- +(-2,-1.5) (v5)-- +(1,0)  (v5) -- +(1,1);
\draw (v6) arc (180:90:.5cm and 3cm);
\draw [dashed] (v0) arc (0:90:.5cm and 3cm);

\draw [gray,fill] (-.8,.5) ellipse (.5cm and 2.5cm) (.8,.5) ellipse (.5cm and 2.5cm);
\end{scope}

\end{tikzpicture} 
\caption{Cutting a surface along a non-separating cycle. The red cycle in the
left diagram is a cycle in the dual.
We cut the surface along the cycle, remove the edges that are cut from
the original graph and attach two topological disks
where the cut is made (right diagram).}
\label{fig:cyclecutting}
\end{figure*}
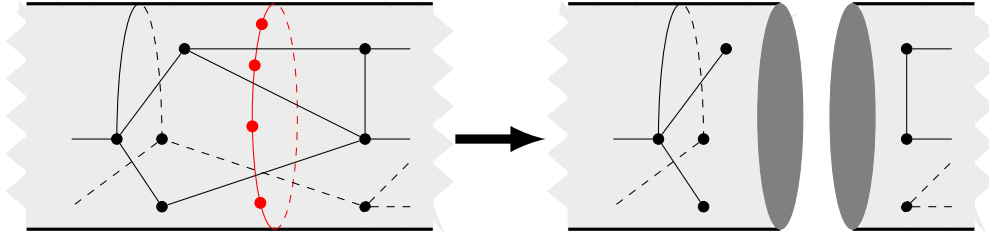

If $C^*$ is a separating cycle, then cutting $\Sigma$ along $C^*$,
creates two surfaces $\Sigma_1$ and $\Sigma_2$, where each one
contains a connected component of $G-C$. The sum of  genera of
$\Sigma_1$ and $\Sigma_2$ is $\genus$. Therefore, in this case only
$\kappa(G-C) > \kappa(G)$.

In both cases the dual embedding is obtained by removing the edges
and vertices of the cycle along which the surface is cut and adding
two vertices $c_l$ and $c_r$ to the disks attached to left and right
side of the cut. The edges of the left side of $C^*$ are attached to $c_l$
and the rest of them to $c_r$.
\end{proof}

Roughly speaking, the  above lemma says that by removing the edges
corresponding to a cycle in $G^*$ from  $G$, we will either decrease
its genus or increase its number of connected components. As we will
show next, the number of connected components in the final graph is
bounded and therefore the procedure has to stop after deleting a
bounded number of cycles.

\begin{proof} [{\bf Lemma \ref{lem:smallconlargegirth}}]
We show that the algorithm \ref{alg:incdualgirth} works correctly.
First of all, note that the algorithm eventually terminates, even if
it has to delete all of the edges of  $G$. Hence, it runs in polynomial
time.

Let $H$ be the output of the algorithm and $H^*$ be its dual. When the algorithm terminates, the girth of $H^*$ is at most $\frac{k}{3\sqrt{\genus}}$. Therefore, the only thing we need to prove is that $\kappa(H) \leq 2\sqrt{\genus}$.

Suppose that the while loop is finished after $m$  iterations.
Consider the total number of edges deleted during the execution of the algorithm. Since the number of deleted edges in each iteration of the while loop is no more than $\frac{k}{3\sqrt{\genus}}$, at most $\frac{mk}{3\sqrt{\genus}}$ edges have been deleted totally. On the other hand, all the edges between different connected components of $H$ have been deleted in the loop. Because $G$ was $k$-edge connected, there was originally $\frac{\kappa(H) k }{2}$ edges between these components, where all of them have been deleted.  Therefore,
\begin{equation}
\label{eq:edgedeletion}
\frac{\kappa(H) k}{2} \leq \frac{mk}{3\sqrt{\genus}}.
\end{equation}

In order to bound $\kappa(H)$, we need an upper bound on $m$. By Lemma \ref{lem:cycleremovalproperty}, we have
\begin{equation*}
\label{eq:mupperbound}
m \leq \left( \genus(G) - \genus(H) \right) + \left(\kappa(H) - \kappa(G)\right)
\leq \genus + \kappa(H).
\end{equation*}

By combining this with inequality (\ref{eq:edgedeletion}) we get

\begin{eqnarray*}
  \frac{\kappa(H)}{2} \leq \frac{\genus+\kappa(H)}{3\sqrt{\genus}}
& \Rightarrow & \frac{\kappa(H)}{6} \leq \frac{\sqrt{\genus}}{3}.
\end{eqnarray*}
This implies that $\kappa(H) \leq 2\sqrt{\genus}.$
\end{proof}




\section{Thin trees, Goddyn's conjecture and ATSP}
\label{sec:final}

The algorithms presented in sections \ref{sec:findthintree} and \ref{sec:dual-girth} and their analysis imply the following result:

\begin{theorem}
\label{thm:constantthintree} A $k$-edge connected graph embeddable
on an orientable surface with genus $\genus$ has a spanning tree
with thinness $\frac{f(\genus)}{k}$ for some function $f(\genus) =
O(\sqrt{\genus}\log(\genus))$. Such a spanning tree can be found in
polynomial time.
\end{theorem}

\begin{proof}
If $\genus = 0$, i.e., $G$ is planar then by \cite{whitney:planargraphs}, $g^*$, the girth of $G^*$ will be at
least $k$.  By Lemma \ref{lem:dual-girth}, Algorithm \ref{alg:computethinset} can find a spanning tree in $G$ with thinness $10/k$.

If $\genus > 0$, then run Algorithm \ref{alg:incdualgirth}  to
obtain a subgraph $H$ of $G$ which has by Lemma
\ref{lem:smallconlargegirth} at most $2\sqrt{\genus}$ connected
components while $girth(H^*) \geq \tfrac{k}{3\sqrt{\genus}}$. Again,
use  Algorithm \ref{alg:computethinset} to find a spanning tree in
each connected component of $H$ with thinness
$\tfrac{6\sqrt{\genus}\mingirth}{k}$ for $\mingirth =4+\lfloor
2\log_2{(\genus+\frac{3}{2})}\rfloor.$ By the matroid property of
spanning trees one can extend this collection of spanning trees to a
spanning tree of $G$ by adding a set $F \subset E(G)$ of size at
most  $2\sqrt{\genus}$. Since $G$ is $k$-edge connected the thinness
 increases by at most $2\sqrt{\genus}/k$. Therefore since $\mingirth \geq 5$ for any $\genus$, the resulting
tree is
$\tfrac{7\sqrt{\genus}\mingirth}{k}=\tfrac{O(\sqrt{\genus}\log{\genus})}{k}$
thin.
\end{proof}

An equivalent way to state the above theorem is that for every
orientable surface $\Sigma$, there exists a function $f_\Sigma$ such
that, for any $\epsilon >0$, every $f_\Sigma(\epsilon)$-edge
connected graph $G$ embedded in $S$ has an $\epsilon$-thin spanning
tree. This can be considered as a partial result for the following
conjecture of Goddyn \cite{goddyn:thintreeconj}.

\begin{conjecture}[Goddyn \cite{goddyn:thintreeconj}]
\label{conj:thintreeconj} There exists a function $f(\epsilon)$ such
that, for any $0 < \epsilon < 1$, every $f(\epsilon)$-edge connected
graph  has an $\epsilon$-thin spanning tree.
\end{conjecture}

Goddyn's conjecture is intimately related to  the asymmetric
traveling salesman problem and the integrality gap of  Held-Karp
relaxation. In order to establish that, we need to extend the definition of thinness
to incorporate costs. We use the following definition due to \cite{atsploglogn} with a slight modification.

\begin{definition} Let $x$ be a feasible  solution of Held-Karp relaxation
(\ref{atsplp}).
We say that a spanning tree $T$ is $\alpha$-thin with respect to $x$,  if for each set $U\subset V$,
$$|T(U,\overline{U})|\leq \alpha x(\delta(U)),$$
where $T(U,\overline{U})$ is the set of the edges of $T$ that are in
the cut $(U, \overline{U})$.  Also we say that $T$  is
$(\alpha,\sigma)$-thin with respect to $x$, if it is $\alpha$-thin
and moreover it is possible to orient the edges of $T$ into $T^*$
(i.e. for each edge $e=(u,v) \in T$, if $c(u,v)<c(v,u)$ add the directed edge $(u,v)$, and otherwise  $(v,u)$ to $T$)
such that
$$c(T^*)\leq \sigma c(x).$$
\end{definition}

Asadpour et al. \cite{atsploglogn} prove the following theorem:

\begin{theorem}[Asadpour et al. \cite{atsploglogn}]
\label{thm:circulation} Assume  that we are given an
$(\alpha,s)$-thin spanning tree $T$ with respect to the LP solution
$x$ of cost $c(x)\leq C\times\HK$. Then we can find a
Hamiltonian cycle of cost no more than $(2\alpha + s)
c(x)=C(2\alpha+s) \HK$ in polynomial time.
\end{theorem}

The above theorem relies on a stronger notion of thinness which takes into consideration the costs of edges.
Proposition \ref{thm:thintreegetridofcost} makes the connection between Goddyn's conjecture and ATSP more concrete by removing the costs.
Let $G(V,E)$ be a weighted undirected graph with weight
function $w(e)$,  and $F \subseteq E$ be a collection of edges. We define
$w(F)  :=  \sum_{e \in F} w(e)$.

\begin{proposition}
\label{thm:thintreegetridofcost} Suppose there exists a
non-decreasing function $g(k)$ such that every $k$-edge connected
graph contains a $\frac{g(k)}{k}$-thin spanning tree. Then every {\em
weighted} $k$-edge connected graph $G(V,E)$ has a $\frac{2g(k)}{k}$-spanning tree $T$ such that $w(T) \leq \frac{2g(k)}{k} w(E)$.
\end{proposition}
\begin{proof}
Let $G_0:=G$ and select a $\frac{g(k)}{k}$-thin spanning tree $T_0$ in $G_0$, and remove its edges. Call this new graph $G_1$. Note that each cut $(U,\overline{U})$ of $G_0$ will lose at most $\frac{g(k)}{k}|G_0(U,\overline{U})|$ of its edges. As the size of the minimum cut in $G_0$ is $k$, $G_1$ will be $(k-g(k))$-edge connected.

Similarly, find a $$\frac{g(k-g(k))}{k-g(k)} \leq \frac{g(k)}{k-g(k)}$$ thin spanning tree $T_1$ in $G_1$. The inequality holds by the monotonicity assumption on $g(k)$. Remove the edges of $T_1$ to obtain a $(k-2g(k))$-edge connected graph $G_2$.
Repeat this algorithm on $G_2$ to obtain $k/2g(k)$ spanning trees $T_0,\ldots,T_ {k/2g(k)-1}$, where for each $i$, $T_i$ is a $\frac{g(k)}{k-i g(k)}$-thin spanning tree of the $(k-ig(k))$-edge connected graph $G_i$.

Because  $G_i$ is a spanning subgraph of $G_0$, any spanning and thin tree of $G_i$ will be spanning and thin in $G_0$. Moreover, since $0\leq i \leq k/2g(k)-1$ and
$$\frac{g(k)}{k-i g(k)} \leq \frac{2g(k)}{k},$$
each $T_i$ is a $\frac{2g(k)}{k}$-thin spanning tree in $G_0$.
Among the selected trees find the one with the smallest weight. Let $T_j$ be that tree. We have
$$\frac{k}{2g(k)} w(T_j) \leq \sum_{i=0}^{k/2g(k)-1} w(T_i) \leq w(G_0).$$

Thus $T_j$ is of the desired thinness and cost.
\end{proof}

Similar to the previous definitions, we will call $T_j$ a $(\frac{2g(k)}{k}, \frac{2g(k)}{k})$-thin spanning tree of $G$.

In the proof of Theorem \ref{thm:constantthintree} we give a
polynomial-time algorithm that finds an
$O(\sqrt{\genus}\log{\genus})/k$-thin spanning tree in a $k$-edge
connected graph embedded on an orientable surface with genus
$\genus$. This result plus the above proposition gives a constant
factor approximation algorithm for ATSP when $\genus$ is constant.
The next theorem establishes this claim.

\begin{theorem}
\label{thm:boundedgenusconstantapp} Given a feasible solution $x$ of
the Held-Karp linear program (\ref{atsplp}),  embedded on an
orientable surface with genus $\genus$,  there is a polynomial-time
algorithm that finds a hamiltonian cycle with a cost that is within
an $O(\sqrt{\genus}\log{\genus})$ of the cost of $x$. In particular,
the approximation factor of the algorithm is at most
$22.5(1+\frac{1}{n})$ when the underlying graph is planar.
\end{theorem}

\begin{proof} 
Let $x$ be a feasible solution of LP (\ref{atsplp})  that can be
embedded on a surface  with genus $\genus$. Construct an undirected
version of $x$ by defining $y_{\{i,j\}} = x_{ij} + x_{ji}$. Define a
new cost function $c'(\{u,v\}) = \min\{c(u,v), c(v,u)\}.$

Round down the fractions in $y$ to the nearest multiple of $1/n^3$.
Construct the integral weighted graph $H$ by adding $n^3
y_{\{i,j\}}$ parallel edges between every pair $i$ and $j$. Since
the size of the support of $y$  is less than $n^2$, we may loose at
most $\frac{1}{n^3}n^2=\frac{1}{n}$ fractions while we are rounding
down the edge fractions and therefore $H$ is
$n^3(2-\frac{1}{n})$-edge connected.

Theorem \ref{thm:constantthintree}  finds a
$\frac{\beta}{n^3(2-\frac{1}{n})}$-thin spanning tree  in $H$, for
$\beta=7\sqrt{\genus}\mingirth$ if $\genus>0$ and $\beta=10$ if $H$ is
planar. Use Proposition \ref{thm:thintreegetridofcost} to compute a
$(\frac{2\beta}{n^3(2-\frac{1}{n})},\frac{2\beta}{n^3(2-\frac{1}{n})})$-thin
tree $T$ in $H$ with respect to cost function $c'$. It is easy to
see that it is possible to orient the edges of $T$ into $T^*$ such
that $c'(T) = c(T^*)$. Since the size of each cut of $H$ is at most $n^3$ times of that  of $x$ and $c(H)\leq n^3c(y)\leq n^3c(x)$, $T^*$ is $(\frac{2\beta}{(2-\frac{1}{n})},\frac{2\beta}{(2-\frac{1}{n})})$-thin with respect to $x$. Therefore, using Theorem
\ref{thm:circulation}, we can find a Hamiltonian cycle of cost no
more than

$$\left(2\frac{2\beta}{2-\frac{1}{n}}+\frac{2\beta}{2-\frac{1}{n}}\right) c(x)\leq 3\beta(1+\frac{1}{n}) c(x)$$

in polynomial time.

Since $\beta=O(\sqrt{\genus}\log{\genus})$ is only a function of
$\genus$, we have a constant factor approximation for ATSP  when the
genus of the graph obtained by $x$ is bounded by a constant. In
particular if $\genus=0$, the above calculation shows that we have a
$30(1+\frac{1}{n})$-approximation algorithm. A slightly better
optimization of parameters  and a minor change of Algorithm
\ref{alg:computethinset}  leads to a $22.5(1+\frac{1}{n})$.

We should also add that Mohar \cite{mohar:embedding} proves for any
constant $\genus$, there is a  linear  time algorithm that finds an
embedding of a given graph on the orientable surface with genus
$\genus$, if such an embedding  exists. Therefore, the embeddability
condition in the above Theorem can be checked in polynomial time for
any constant $\genus$.
\end{proof}

\begin{remark}
It is worth noting that the genus of an extreme point solution of
Held-Karp relaxation instance with $n$ vertices can be as large as
$\Omega(n)$. In fact, for any odd $r$, it is possible to construct an
extreme point on $r^2$ vertices  that has $K_r$ as a minor. Such an
extreme point can be obtained by the same construction as Carr  and
Vempala \cite[Theorem 3.5]{vempala:4/3integralitygap}  applied to
$K_r$.
\end{remark}

An argument similar to the proof of Theorem
\ref{thm:boundedgenusconstantapp} shows that Goddyn's conjecture
implies constant integrality gap of the Held-Karp relaxation of
ATSP. Furthermore, an algorithmic proof of Goddyn's conjecture
implies a constant factor approximation algorithm for ATSP.

\begin{corollary}
\label{cor:goddynconstintegatsp} If Goddyn's conjecture is true for
some function $f(\epsilon) = O(1/\epsilon)$, then the integrality
gap of Held-Karp relaxation is bounded from above by a constant.
\end{corollary}

\subsection{Nowhere-zero flows and Jaeger's conjecture}
\label{sec:goddyn}

Goddyn's conjecture was inspired by the study of nowhere-zero flows
and in particular in attempting Jaeger's
conjecture~\cite{jaeger:flowconj}. Here, we just state the Jaeger's
conjecture  and refer the reader to Seymour
\cite{seymour:handbookcombinatorics} for more information.

\begin{conjecture}[Jaeger \cite{jaeger:flowconj}]
\label{conj:circularflow} If $G$ is $4k$-edge connected, then for
some orientation of $G$, every subset $S \subset V(G)$ satisfies
$$(k-1) |\delta^-(S)| \leq k |\delta^+(S)| \leq (k+1) |\delta^-(S)|.$$
\end{conjecture}

Jaeger's conjecture has not been proved for any positive integer
$k$ yet. For $k=1$, this is Tutte's 3-flow conjecture and is proved only
for planar graphs. Goddyn's conjecture implies a weaker version of
Jaeger's conjecture in which $4k$ is replaced by an arbitrary
function of $k$.
Even this version is
still open \cite{goddyn:largegirth}.
%
%

Previously, Zhang \cite{zhang:jaeger}  proved Jaeger's conjecture on
graphs with bounded genus. The dependence of his result on the genus
of the graph is quite similar to ours: he proves Jaeger's
conjecture for graphs with connectivity
$O(\sqrt{\genus}k)$ embedded on a surface with genus $\genus$. Since
Goddyn's conjecture implies Jaeger's conjecture with the same
parameters, our result can be seen as a strengthening of
\cite{zhang:jaeger} for surfaces with orientable genus. We leave the
argument for surfaces with bounded non-orientable genus as an open problem.


As a final note, it is easy to extend the polynomial-time dynamic programming algorithm for solving TSP on graphs with bounded treewidth to ATSP. Nevertheless,  we still do not know if every $k$-edge connected graph with  bounded treewidth has an $O(1)/k$-thin tree. The answer to this question will be helpful in finding thin trees in families of graphs with excluded minors. \\

{\bf Acknowledgements.}  The authors would like to thank Arash
Asadpour, Luis Goddyn, Michel Goemans, Sergey Norin, Dana Randall,
David Shmoys, Prasad Tetali, and Jan Vondrak for discussions and
comments on an earlier version of this paper. The authors were also supported by
the National Science Foundation and Stanford Graduate Fellowship.
\bibliographystyle{abbrv}
\bibliography{references}

\begin{thebibliography}{10}

\bibitem{aroraeuclidean}
S.~Arora.
\newblock Polynomial time approximation schemes for euclidean traveling
  salesman and other geometric problems.
\newblock In {\em Journal of the ACM}, pages 2--11, 1996.

\bibitem{Arora:planartsp}
S.~Arora, M.~Grigni, D.~Karger, P.~Klein, and A.~Woloszyn.
\newblock A polynomial-time approximation scheme for weighted planar graph tsp.
\newblock In {\em SODA '98: Proceedings of the ninth annual ACM-SIAM symposium
  on Discrete algorithms}, pages 33--41, Philadelphia, PA, USA, 1998. Society
  for Industrial and Applied Mathematics.

\bibitem{atsploglogn}
A.~Asadpour, M.~X. Goemans, A.~Madry, S.~Oveis~Gharan, and A.~Saberi.
\newblock O(logn/loglogn) approximation to the asymmetric traveling salesman
  problem.
\newblock {\em to appear in SODA 2010}, 2010.

\bibitem{bgrs04}
V.~Bil{\`o}, V.~Goyal, R.~Ravi, and M.~Singh.
\newblock On the crossing spanning tree problem.
\newblock In {\em APPROX-RANDOM}, pages 51--60, 2004.

\bibitem{Blaser02}
M.~{Bl\"aser}.
\newblock A new approximation algorithm for the asymmetric {TSP} with triangle
  inequality.
\newblock In {\em SODA}, pages 638--645, 2002.

\bibitem{vempala:4/3integralitygap}
R.~D. Carr and S.~Vempala.
\newblock Towards a 4/3 approximation for the asymmetric traveling salesman
  problem.
\newblock In {\em SODA '00: Proceedings of the eleventh annual ACM-SIAM
  symposium on Discrete algorithms}, pages 116--125, Philadelphia, PA, USA,
  2000. Society for Industrial and Applied Mathematics.

\bibitem{charikar:integralitygap}
M.~Charikar, M.~X. Goemans, and H.~Karloff.
\newblock On the integrality ratio for the asymmetric traveling salesman
  problem.
\newblock {\em Math. Oper. Res.}, 31(2):245--252, 2006.

\bibitem{c10}
C.~Chekuri.
\newblock private communication.
\newblock 2010.

\bibitem{cvz09}
C.~Chekuri, J.~Vondr{\'a}k, and R.~Zenklusen.
\newblock Dependent randomized rounding for matroid polytopes and applications.
\newblock {\em CoRR}, abs/0909.4348, 2009.

\bibitem{Christofides76}
N.~Christofides.
\newblock Worst case analysis of a new heuristic for the traveling salesman
  problem.
\newblock Report 388, Graduate School of Industrial Administration,
  Carnegie-Mellon University, Pittsburgh, PA, 1976.

\bibitem{Demaine:boundedgenus}
E.~D. Demaine, M.~Hajiaghayi, and B.~Mohar.
\newblock Approximation algorithms via contraction decomposition.
\newblock In {\em SODA '07: Proceedings of the eighteenth annual ACM-SIAM
  symposium on Discrete algorithms}, pages 278--287, Philadelphia, PA, USA,
  2007. Society for Industrial and Applied Mathematics.

\bibitem{FeigeS07}
U.~Feige and M.~Singh.
\newblock Improved approximation ratios for traveling salesman tours and path
  sin directed graphs.
\newblock In {\em APPROX}, pages 104--118, 2007.

\bibitem{flm04}
S.~P. Fekete, M.~E. L\"{u}bbecke, and H.~Meijer.
\newblock Minimizing the stabbing number of matchings, trees, and
  triangulations.
\newblock In {\em SODA '04: Proceedings of the fifteenth annual ACM-SIAM
  symposium on Discrete algorithms}, pages 437--446, Philadelphia, PA, USA,
  2004. Society for Industrial and Applied Mathematics.

\bibitem{atsplogn}
A.~M. Frieze, G.~Galbiati, and F.~Maffioli.
\newblock On the worst-case performance of some algorithms for the asymmetric
  traveling salesman problem.
\newblock {\em Networks}, 12:23--39, 1982.

\bibitem{goddyn:thintreeconj}
L.~A. Goddyn.
\newblock Some open problems i like.
\newblock available at
  {\url{http://www.math.sfu.ca/~goddyn/Problems/problems.html}}.

\bibitem{goddyn:largegirth}
L.~A. Goddyn, P.~Hell, A.~Galluccio, and A.~Galluccio.
\newblock High girth graphs avoiding a minor are nearly bipartite.
\newblock {\em J. Comb. Theory (B)}, 83:1--14, 1999.

\bibitem{goemans06}
M.~X. Goemans.
\newblock Minimum bounded degree spanning trees.
\newblock In {\em FOCS}, pages 273--282, 2006.

\bibitem{Papadimitriou:planartsp}
M.~Grigni, E.~Koutsoupias, and C.~Papadimitriou.
\newblock An approximation scheme for planar graph tsp.
\newblock In {\em FOCS '95: Proceedings of the 36th Annual Symposium on
  Foundations of Computer Science}, page 640, Washington, DC, USA, 1995. IEEE
  Computer Society.

\bibitem{Kaplan05}
N.~S. H.~Kaplan, M.~Lewenstein and M.~Sviridenko.
\newblock Approximation algorithms for asymmetric {TSP} by decomposing directed
  regular multigraphs.
\newblock {\em J. {ACM}}, pages 602--626, 2005.

\bibitem{h09}
S.~Har-Peled.
\newblock Approximating spanning trees with low crossing number.
\newblock Technical report, 2009.

\bibitem{HeldKarp70}
M.~Held and R.~Karp.
\newblock The traveling salesman problem and minimum spanning trees.
\newblock {\em Operations Research}, 18:1138--1162, 1970.

\bibitem{jaeger:flowconj}
F.~Jaeger.
\newblock On circular flows in graphs, finite and infinite sets.
\newblock {\em Colloquia Mathematica Societatis Janos Bolyai 37}, pages
  391--402, 1984.

\bibitem{kleintsp}
P.~N. Klein.
\newblock A linear-time approximation scheme for planar weighted tsp.
\newblock In {\em FOCS '05: Proceedings of the 46th Annual IEEE Symposium on
  Foundations of Computer Science}, pages 647--657, Washington, DC, USA, 2005.
  IEEE Computer Society.

\bibitem{mohar:embedding}
B.~Mohar.
\newblock Embedding graphs in an arbitrary surface in linear time.
\newblock In {\em STOC '96: Proceedings of the twenty-eighth annual ACM
  symposium on Theory of computing}, pages 392--397, New York, NY, USA, 1996.
  ACM.

\bibitem{moharthomassen}
B.~Mohar and C.~Thomassen.
\newblock {\em Graphs on Surfaces}.
\newblock Johns Hopkins University Press, Baltimore, 2001.

\bibitem{seymour:handbookcombinatorics}
P.~D. Seymour.
\newblock Nowhere zero flows.
\newblock In R.~L. Graham, M.~Gr\"{o}tschel, and L.~Lov\'{a}sz, editors, {\em
  Handbook of combinatorics}, volume~1. MIT Press, Cambridge, MA, USA, 1995.

\bibitem{whitney:planargraphs}
H.~Whitney.
\newblock Non-separable and planar graphs.
\newblock {\em Trans. Amer. Math. Soc}, 34(2):339--362, 1932.

\bibitem{zhang:jaeger}
C.~Q. Zhang.
\newblock Circular flows of nearly eulerian graphs and vertex-splitting.
\newblock {\em J. Graph Theory}, 40(3):147--161, 2002.

\end{thebibliography}

\appendix
\section{A Constant Factor Approximation Algorithm for the Minimum Stabbing Tree Problem}
\label{sec:stabbingtree}

Another interesting application of Theorem \ref{thm:constantthintree} is to the minimum stabbing tree problem. 
The minimum stabbing tree problem arises in computational geometry: the input is a set $P ={p_1, \ldots, p_n}$ of points in $R^d$. The task is to construct a spanning tree on $P$ by connecting vertices with straight lines such that the crossing number, which  is the maximum number of segments that are encountered (in their interior or at an endpoint) by any line, is minimized. The problem is known to be NP-hard by the work of Fekete et al \cite{flm04}. Bilo et al. \cite{bgrs04} and HarPeled \cite{h09} in two disjoint works found an $O(\log{n})$ approximation algorithm for the problem in $d$-dimensional space. More recently, Chekuri et al. \cite{cvz09} derive an $O(\log n/\log\log n)$ approximation algorithm for these problems

Fekete et al. in their work \cite{flm04} also considered the natural Linear Programming relaxation of the problem and they proved it contains a fractional optimal solution with planar support graph:

\begin{theorem}[Fekete et al. \cite{flm04}]
\label{thm:stabbing_planarsupport}
For any set of $P$ vertices in the {\em plane}, there is a fractional spanning tree $x^*$ of minimum stabbing number such that the support graph of $x$ is planar. Such a fractional spanning tree can be found in polynomial time.
\end{theorem}

Using this theorem Fekete  et al. show that there is a fractional spanning tree $x^*$ of minimum stabbing number that has an edge of weight more than $1/3$. However, they left it as an open problem to find a spanning tree with a constant-factor guarantee. 


We can use
Theorem \ref{thm:constantthintree} to obtain a constant factor approximation algorithm for the minimum stabbing tree problem  for $d=2$.
\begin{corollary}
\label{cor:stabbing_constantfactor}
Let $P$ be a set of $n$ points in the plane. There is a deterministic polynomial time algorithm that finds a spanning tree $T$ with crossing number at most $10(1+o(1))$ times the minimum crossing number of any spanning tree of $P$.
\end{corollary}
\begin{proof}
By Theorem \ref{thm:stabbing_planarsupport}, we can find a fractional spanning tree $x^*$ with minimum stabbing number and planar support in polynomial time. Similar to the proof of Theorem \ref{thm:boundedgenusconstantapp}, we can round down $x$ to the nearest multiple of $1/n^3$ and then construct a graph $G$ by adding $n^3 x_{i,j}$ parallel edges between every pair $i$ and $j$. Similarly, $H$ will be $n^3(1-\frac{1}{n})$-edge connected. By Theorem \ref{thm:constantthintree}, we can find a $\frac{10}{n^3(1-1/n)}$-thin spanning tree $T$ in $G$ in polynomial time. The number of edges of $T$ across any cut in $G$ (including the ones  corresponding to any line in the plane) is at most $10\frac{n^3}{n^3(1-1/n)}=10(1+o(1))$ times the size of that cut in $x^*$. Therefore, the crossing number of $T$ is at most a constant times the minimum crossing number of any spanning tree of $P$.
\end{proof}

\end{document}